\documentstyle[prb,aps,graphicx,multicol]{revtex}

\begin{document}
\draft
\title{Correlation gap in the heavy-fermion antiferromagnet UPd$_2$Al$_3$}
\author{M. Dressel\cite{email}, N. Kasper\cite{permanent2}, K. Petukhov, D.N. Peligrad, and B. Gorshunov\cite{permanent1}}
\address{Physikalisches Institut, Universit\"at Stuttgart,
Pfaffenwaldring 57, D-70550 Stuttgart, Germany}
\author{M. Jourdan, M. Huth, and H. Adrian}
\address{Institut f\"ur Physik, Universit\"at Mainz,
Staudinger Weg 7, D-55099 Mainz, Germany}
\date{Received  \today}
\maketitle

\begin{abstract}
The optical properties of the heavy-fermion compound
UPd$_2$Al$_3$ have been measured in
the frequency range from 0.04~meV to 5~meV (0.3 to 40~cm$^{-1}$) at
temperatures $2~{\rm K}<T< 300$~K. Below the coherence temperature
$T^*\approx 50$~K, the hybridization gap
opens around 10~meV. As
the temperature decreases further ($T\leq 20$~K),
a well pronounced pseudogap
of approximately $0.2$~meV develops in the optical response;
we relate this to the antiferromagnetic ordering which occurs below 
$T_N\approx 14$~K.
The frequency dependent mass and scattering rate give evidence 
that the enhancement of the effective mass mainly
occurs below the energy which is associated to the magnetic correlations 
between the itinerant and localized $5f$
electrons. In addition to this correlation gap, we observe a narrow  
zero-frequency conductivity peak which at
2~K is less than 0.1~meV wide, and which contains only a fraction of 
the delocalized carriers. The analysis
of the spectral weight infers a loss of kinetic energy associated 
with the superconducting transition.
\end{abstract}


\begin{multicols}{2}
\columnseprule 0pt
\narrowtext

\section{Introduction}
More than twenty years after the discovery of superconductivity 
in heavy-fermion (HF) compounds,\cite{Steglich79}
the nature of the superconducting ground state is still under debate. 
In particular the discovery of the
coexistence of antiferromagnetic (AF) ordering and superconductivity 
in some of these materials has reinforced
the interest to study the interplay between electronic and magnetic degrees 
of freedom in these correlated
electron systems. Instead of the conventional electron-phonon interaction, 
the electron-magnon correlations have
been proposed to cause the pairing in HF superconductors.\cite{Ott87,Grewe91} 
Heavy fermions are intermetallic
compounds containing elements with $f$ electrons which show a large 
enhancement of the quasiparticle effective
mass. The wavefunctions of the atomic and conduction electrons 
mix (hybridization), and the strong interaction of the quasi-free
conduction-band electrons with nearly localized $f$ electrons leads to a so-called Abrikosov-Suhl resonance, i.e.\ an enhanced 
density of states at the Fermi energy. \cite{Grewe91} 
The main idea of understanding this state of matter is the competition 
between Kondo and
Ruderman-Kittel-Kasuya-Yosida (RKKY) interactions. The RKKY interaction leads 
to a magnetic ground state while the
dominance of the Kondo interaction causes hybridization between localized 
$f$-electron states and delocalized
conduction electrons. Due to the arrangement of the $f$ orbitals
on a regular lattice,  
the hybridization with the itinerant states leads to the opening of a charge gap 
at the Fermi energy; hence if not
fully developed at least a pseudogap should show up in the electronic excitation
spectrum.\cite{Millis87,Coleman87,Georges96,Fulde93,Hewson97}

The common feature of the HF systems is the
crossover from an incoherent state, where the scattering of
the charge carriers
on magnetic moments can be described in the frame of the
single-particle Kondo model, to a many-body ground state (Kondo
lattice) on cooling below the coherence temperature $T^{*}$.
Above $T^{*}$ the optical conductivity exhibits a broad Drude
behavior
\begin{equation}
\hat{\sigma}(\omega)=\sigma_1(\omega)+i\sigma_2(\omega)=
\frac{\sigma_{dc}}{1+i\omega\tau}
\label{eq:metalDrude}
\end{equation}
which characterizes normal metals;\cite{DresselGruner} the dc conductivity 
is given by
$\sigma_{dc}=ne^2\tau/m$
with $n$ the concentration of conduction electrons, $e$ and $m$ the electronic 
charge and mass, and $1/\tau=\Gamma$ the scattering rate, which is typically a 
few hundred wavenumbers for usual metals.
The area under the conductivity spectrum is the spectral weight:
\begin{equation}
\int  \sigma_1(\omega)\,{\rm d}\omega = \frac{\pi n e^2}{2 m} = 
\frac{\omega _p^{2}}{8}
\label{eq:metalsumrule}
\end{equation}
and determines the plasma frequency $\omega_p=\sqrt{4\pi n e^2/m}$.

Below the coherence temperature $T^{*}$, however,
the increase in the density of electronic states (DOS) at low energies 
is commonly described by the formation of  a  narrow
Drude peak with a renormalized effective mass $m^{*}$  and
scattering rate $\Gamma^*$  of charge 
carriers:\cite{Fukuyama85,Wachter93,Degiorgi99}
\begin{equation}
\hat{\sigma}(\omega)= \frac{(\omega_p^{*})^2}{4\pi}
\frac{1}{\Gamma^{*}+i\omega} .
\label{eq:renDrude}
\end{equation}
Because $m^*\Gamma^*=m\Gamma$, the dc conductivity
$\sigma_{dc}=ne^2/(m^*\Gamma^*)$ is not affected by the renormalization.
The spectral weight $\omega _p^{* 2}/8=\pi n e^2/(2 m^{*})$,
however,  decreases as $m^*$ increases; typically $m^*$ can reach several
hundred times the free electron mass $m$.
The enhancement of the effective mass  is connected
to the hybridization of $s$ and $f$-orbitals which leads to a narrow
band with a high DOS at the Fermi energy
 and to the strong coupling of the charge carriers to the
local magnetic moments.\cite{Fukuyama85,Wachter93}
In HF compounds which do not order magnetically, a scaling relation
\begin{equation}
\left (\frac{\Delta}{T^*}\right)^2 = \frac{m^*}{m}
\label{eq:scaling}
\end{equation}
between the hybridization gap $\Delta$ and the enhancement of the effective mass 
$m^*/m$ is found theoretically \cite{Millis87,Coleman87,Georges96,Fulde93,Hewson97}
and experimentally.\cite{Marabelli90,Bocelli94,Dordevic01} With $T^*$ in the order 
of 10 to 100~K the hybridization gap is normally observed 
in the far-infrared range of frequency.

The antiferromagnetic ordering reveals itself in rather different ways for the various 
HF compounds.\cite{Degiorgi99,Degiorgi97} For instance, UPt$_3$ does not show\cite{Stewart84,Donovan97} any
appreciable anomaly of the specific heat or resistivity at the N{\'e}el 
temperature $T_N=5$~K while in
URu$_2$Si$_2$ and UPd$_2$Al$_3$ the transition into a magnetically ordered 
state is accompanied by clear
anomalies of these quantities.\cite{Maple86,Geibel91} 
However, URu$_2$Si$_2$ shows a remarkable resistivity
increase at $T_N$, typical for a spin-density-wave (SDW) system, 
while in UPd$_2$Al$_3$ only a kink in the
$\rho(T)$ due to freezing-out of the spin-flip-scattering is observed.\cite{Degiorgi97} 
This already indicates
that in contrast to the former system with itinerant antiferromagnetism, 
UPd$_2$Al$_3$ is an antiferromagnet with
localized spins. In both UPt$_3$ and URu$_2$Si$_2$, but also in UCu$_5$, 
the progressive development of a gap (or pseudogap) in the optical spectra is 
observed in connection with the ordering of the magnetic
moments,\cite{Degiorgi97,Donovan97,Bonn88,Degiorgi94b} 
whereas for UPd$_2$Al$_3$ no effect of magnetic ordering
on the optical response at frequencies above 30~cm$^{-1}$ 
has been detected in the first infrared
investigations.\cite{Degiorgi97,Degiorgi94}

In the heavy fermion superconductor UPd$_2$Al$_3$,
on which we want to focus in this paper, a maximum 
of the resistivity occurs at  temperaturea around 80~K 
which is often related to the onset of coherence;\cite{remark5}
the values cited for the coherence temperature $T^*$ range from 20~K to
60 K depending on the experimental method used for determination.
The magnetic susceptibility $\chi (T)$ shows  
appreciably different behaviors for field directions parallel and
perpendicular to the hexagonal $c$-axis;\cite{remark6} in the 
$ab$-plane $\chi (T)$ is Curie-Weiss like above 40~K with a
maximum at around 35~K. Below this temperature,  AF correlations develop, 
and  UPd$_2$Al$_3$ shows a metamagnetic
behavior. When cooling down even further, a commensurate AF order develops 
below $T_N=14$~K; superconductivity
finally sets in\cite{Geibel91,Paolasini93,Geibel912} below 2~K. 
For $2~{\rm K}<T<14$~K 
the specific heat shows a $C/T\propto T^2$ dependence; the
effective mass of  the charge carriers is estimated\cite{Geibel91,Dalichaouch92} 
as $m^*/m \approx 50$.

In the superconducting state of UPd$_2$Al$_3$ 
two  low-energy modes at 1.5~meV and 0.4~meV
have been observed\cite{Bernhoeft98}
by inelastic neutron scattering experiments at ${\bf q}=\frac{1}{2c}(0, 0, 1)$; 
it was suggested that they are associated
with magnetic excitons and superconductivity, respectively.
An important feature of UPd$_2$Al$_3$ is that --~compared with
other HF antiferromagnets~-- it has a rather large magnetic moment
(0.85 $\mu_{B}$)
which is localized predominantly on the
U-site.\cite{Paolasini93,Krimmel92} Based on this experimental evidence, it
has been suggested that UPd$_2$Al$_3$ can be described as a
local-moment magnet, and hence the magnetic ordering should have
only a minor influence on the electronic DOS. However, theoretical
calculations of the Fermi  surface \cite{Sandratskii94} as well
as de Haas-van Alphen experiments \cite{Inada94} also indicate an
itinerant character of $5f$-electrons, and therefore one can
expect in UPd$_2$Al$_3$ to find appreciable correlations between
electronic states and magnetic ordering.
Besides the interesting interplay of magnetic ordering and superconductivity,
these correlations make UPd$_2$Al$_3$  one of the most studied HF 
systems in recent years; they are the main subject of our investigation.

Optical experiments have in general proven to be sensitive to the formation
of heavy quasiparticles;\cite{Wachter93,Degiorgi99}
the relevant features show up at the lower end of the infrared spectral range
which is commonly accessible.
In order to extend our earlier investigations \cite{Degiorgi97,Degiorgi94} to lower
frequencies, in the present work
we have carried out optical experiments on UPd$_2$Al$_3$ films
in the spectral range from microwaves up to the far infrared.
First results have partially
been reported   in Refs.~\onlinecite{Dressel98,Dressel00,Dressel02}.

\section {Experimental Methods}
The highly $c$-axis oriented, epitaxial  thin ($150$~nm) film of UPd$_2$Al$_3$ 
was prepared on (111) oriented LaAlO$_3$
substrate (thickness $0.924$~mm) by electron-beam co-evaporation 
of the constituent elements in a molecular-beam
epitaxy system.\cite{Huth93} The phase purity and structure of the film 
were investigated by X-ray and reflection
high-energy electron diffraction. The high quality of the film is seen in 
dc resistivity data displayed in
Fig.~\ref{fig:figure1} which  are in excellent agreement with measurements of bulk crystals.\cite{Geibel91} 
At $T=300$~K the resistivity is $\rho=172~\mu\Omega$cm, 
the residual resistivity ratio $\rho_{300\rm K}/\rho_{2\rm
K} = 23$, the N{\'e}el temperature $T_N=14$~K; 
our film shows a sharp superconducting transition at $T_c=1.75$~K.

For the measurements in the millimeter and sub-millimeter spectral range 
($50-1400$~GHz, $1.2-40$~cm$^{-1}$,
$0.2 - 5$~meV) a coherent source spectrometer was
employed \cite{Volkov85} utilizing a set of different backward wave 
oscillators as monochromatic and continuously
tunable sources. The spectrometer includes an interferometric setup 
in a Mach-Zehnder arrangement which allows
for measuring both the amplitude $T_F$ and the phase $\phi_t$ 
of the signal transmitted through the plane-parallel
sample, which in our case is a film on a substrate. 
These two quantities are used to evaluate the complex
refractive index $\hat{N}=n+ik$ [or alternatively the complex conductivity $\hat{\sigma}(\omega)$] of the film
using Fresnel's formulas for a two-layer system \cite{DresselGruner} 
without assuming any particular model. In
general, the complex transmission coefficient $\hat{t}_{1234}=\sqrt{T_F}\, 
e^{i\phi_t}$ of a two-layer system
(layers which we index with 2 and 3) separating two media (indices 1 and 4) is given by
\begin{eqnarray}
\hat{t}_{1234} = \hat{t}_{12} \hat{t}_{23} \hat{t}_{34} \exp
\bigl\{i(\delta_2 + \delta_3)\bigr\} \left[ 1 +  \hat{r}_{12} \hat{r}_{23} \exp \bigl\{2i\delta_2 \bigr\}\right.\nonumber\\
\left. +  \hat{r}_{23} \hat{r}_{34}
\exp\bigl\{2i\delta_3 \bigr\} +
\hat{r}_{12} \hat{r}_{34} \exp \bigl\{2i (\delta_2 + \delta_3) \bigr\}\right]^{-1}
\end{eqnarray}
where the complex angles are $\delta_2 = \omega d_{2}(n_2 + i k_2)/c$ 
for the film and $\delta_3 = \omega
d_{3}(n_3 + i k_3)/c$ for the substrate; $c=3.0 \cdot 10^{10}$~cm/s 
is the velocity of light in vacuum. 
The transmission coefficients at each interface are evaluated by the
standard equation $\hat{t}_{ij} = (\hat{N}_i - \hat{N}_j)(\hat{N}_i  + \hat{N}_j)^{-1}$; where
$\hat{N}_1=\hat{N}_4=1$ correspond to vacuum. 
The optical parameters of the LaAlO$_3$ substrate ($n_3$ and $k_3$)
are determined beforehand by performing the experiments on a blank substrate. 
The large size of the sample
(approximately $10~ \times 10$~mm$^2$) allowed us to extend the 
measurements to very low frequencies, from
$\omega/(2\pi c)= 40$~cm$^{-1}$ down to 1.15~cm$^{-1}$.

For the infrared range and higher frequencies we used the reflectivity results 
obtained on bulk samples and published
previously.\cite{Degiorgi97,Degiorgi94}

In addition, at the frequencies of 10, 24, and 34~GHz enclosed
resonators were developed and the sample was measured by
cavity perturbation technique.\cite{Klein93c,Donovan93c,Dressel93c,Peligrad98}
The cylindrical TE$_{011}$ copper cavities --~quality factors
around 15\,000~-- were operated by
Gunn diode oscillators with sufficient tuning range;
two waveguides couple to the cavity by
holes on the half height of the opposite side walls.
A small slice of a sample (typically 3 $\times$ 3 mm$^2$)
was placed in the electrical field maximum. The
transmitted power was detected by a diode, amplified and
processed by fitting a Lorentzian in order to determine the central
frequency $f_0$ and the width $\Gamma_0$.
From the shift in frequency $\Delta f$ and change in
width $\Delta\Gamma$ upon introduction of the sample,
the surface resistance $R_S$ and the surface reactance $X_S$ 
which determine the
surface impedance $\hat{Z}_S=R_S+iX_S$ are calculated as follows:
\begin{equation}
R_S=Z_0\frac{\Delta\Gamma}{2f_0\zeta} \hspace*{0.5cm} {\rm and}
\hspace*{.5cm} X_S=Z_0\frac{\Delta f}{f_0\zeta},
\end{equation}
where 
$Z_0={4\pi}/{c}=4.19\cdot  10^{-10}$~s/cm is the impedance of free
space ($Z_0=377~\Omega$ in SI units). The resonator constant $\zeta$ can
be calculated from the geometry of the cavity and the sample
when the dielectric properties of
the substrate are known.\cite{Klein93c,Peligrad98}
The complex conductivity $\hat{\sigma}$ can be evaluated
from the surface impedance using $\hat{Z}_S=Z_0\sqrt{\omega/4\pi i\hat{\sigma}}$,
and the absorptivity $A$ (sometimes simply called the absorption) is given by the relation
\begin{equation}
A =1-R = \frac{4R_S}{Z_0}\left(1+\frac{2R_S}{Z_0}+
\frac{R_S^2+X_S^2}{Z_0^2}\right)^{-1},
\label{eq:mw-abs}
\end{equation}
where $R$ is the reflectivity. In the limit  $R_S, |X_S| \ll
Z_0$, the last equation reduces to
\begin{equation}
A\approx {4R_S}/{Z_0}.
\label{eq:mw-abs-hr}
\end{equation}

While at 10 GHz the two components of the complex surface impedance could be measured and therefore both the conductivity and the dielectric constant 
were calculated,
in our higher frequency experiments at 24 and 34~GHz 
--~because of the large influence of  the substrate~-- 
we were not able to determine the frequency shift accurately enough
to evaluate the dielectric properties of the film; hence we only utilize 
the surface resistance $R_S$.
By using Eq.~(\ref{eq:mw-abs-hr}), however, it was possible to 
combine these millimeter wave cavity data with all of the higher
frequency optical data, i.e.\ the transmission results through the films 
obtained in the submillimeter wave range and the 
reflectivity \cite{Degiorgi97,Degiorgi94}
measured on bulk samples in the
infrared, visible and ultraviolet spectral range.
The reflection coefficient is a complex
function $\hat{r}(\omega)=|r(\omega)|e^{i\phi_r(\omega)}$ with the
measured bulk reflectivity $R=|\hat{r}|^2$.
In order to obtain the phase $\phi_r$ we performed a
Kramers-Kronig analysis on the reflectivity spectra: \cite{DresselGruner}
\begin{equation}
\phi_r(\omega) =
\frac{\omega}{\pi}\int_{0}^{\infty}\frac{\ln{[R(\omega')]} -
\ln{[R(\omega)]}}{\omega^2 - (\omega')^2}\,{\rm d}\omega'\, ,
\label{KK}
\end{equation}
where the $\ln{[R(\omega)]}$ term has been added to the standard form in
order to remove the singularity at $\omega'=\omega$. It has no effect on the
integral because $\int_0^\infty [\omega^2 - (\omega')^2]^{-1}{\rm d}\omega'=0$.
Because this
integral extends from zero to infinity, it is necessary to make suitable
high and low frequency extrapolations to the measured reflectivity data.
We have chosen to use a power law [$R(\omega)\propto
1/\omega^4$] at high frequencies  above $10^6$~cm$^{-1}$
and a Hagen-Rubens extrapolation 
\begin{equation}
R(\omega)=1-\left(\frac{2\omega}{\pi\sigma_{\rm dc}}\right)^{1/2} 
\label{eq:hr}
\end{equation}
below 0.3~cm$^{-1}$.  From $R(\omega)$ and $\phi_r(\omega)$
it is then possible\cite{DresselGruner} to calculate the complex optical
conductivity $\hat{\sigma}(\omega)$.

\section{Results and Analysis}

\subsection{Transmission Spectra}
\label{sec:transmissionspectrum} 
In Fig.~\ref{fig:figure2} we present the low-frequency transmission spectra for
the UPd$_2$Al$_3$ film on the substrate for two selected temperatures 
(100~K and 2~K); they already illustrate
clearly our main finding. The fringes in the spectra are due to multi-reflection 
of the radiation within the
plane-parallel substrate acting as a Fabry-Perot re\-sonator 
for our monochromatic radiation.\cite{DresselGruner}
The frequency spacing between two peaks is mainly determined 
by the thickness and refractive index of the
substrate; their amplitude (minima to maxima), however, is governed 
by the parameters of the film. At
temperatures above 25~K, the transmission 
spectra basically do not change. 
For $T\leq 20$~K the overall transmission is found to be 
strongly reduced below $10~{\rm cm}^{-1}$ due
to absorption within the film. At lower frequencies, 
$\nu=\omega/(2\pi c) <3.5~{\rm cm}^{-1}$, the transmitted
signal increases again, indicating an absorption edge. 
It is worth to note, that our preliminary study of a
136~nm thick UPd$_2$Al$_3$ film  showed the same 
features in the optical response.\cite{Dressel98}

This feature appears below the temperature where the magnetic 
susceptibility has a maximum --~when the AF
correlations develop. Therefore we checked whether the observed
transmission minimum could be due to some magnetic absorption.
We tried to fit the observed 
behavior by modelling the  complex permeability $\hat{\mu}(\omega)$ 
with a magnetic oscillator (Lorentzian):
\begin{equation}
\hat{\mu}(\omega) = 1 +
\frac{\Delta\mu_1 \omega_0^2}{\omega_0^2 - \omega^2 + i\omega \Gamma}
\end{equation}
where $\omega_0$ and $\Gamma$ are the center frequency and the 
width of the oscillator, respectively, and $\Delta
\mu\,\omega_0^2$ denotes the spectral weight. 
Alternatively we can satisfactorily describe the transmission 
spectra by using a dielectric
oscillator
\begin{equation}
\hat{\epsilon}(\omega) =  \epsilon_1(\omega) + i\epsilon_2(\omega) = 
1 + \frac{\Delta\epsilon_1\omega_{0}^2}{\omega_0^2-\omega^2+
i\omega\Gamma}.
\end{equation}
Here $\Delta\epsilon_1\omega_0^2$ describes the oscillator strength (spectral weight),
$\omega_0$ and $\Gamma$ denote the center frequency and the width of the harmonic oscillator.
The results of the least-square fit of the transmitted signal using a 
Drude--Lorentz  model in combination with a
magnetic or with a dielectric oscillator are shown in the lower 
two panels of Fig.~\ref{fig:figure2}. The
parameters of the magnetic oscillator are $\Delta \mu_1=1.001$, 
$\omega_0/(2 \pi c)=5.45$~cm$^{-1}$, $\Gamma/(2
\pi c)=4.08$~cm$^{-1}$; and for the dielectric oscillator we used 
accordingly $\Delta
\epsilon_1=\omega_p^2/\omega_0^2= 2.14\cdot 10^5$, 
$\omega_0/(2 \pi c)= 3.86$~cm$^{-1}$, $\Gamma/(2 \pi c)=2.13$~cm$^{-1}$. 
One can  immediately see that, 
no satisfactory fit is possible with a magnetic oscillator term.
Second, the $\Delta \mu_1$ value 
(which denotes the strength of the 
magnetic contribution) is much higher than expected from the static magnetic 
susceptibility measurements.\cite{Geibel91}
Third, the frequency dependence of the transmission is only described 
satisfactorily using the  dielectric
model. Thus we conclude that the observed feature is not due to a pure 
magnetic excitation and proceed with the
analysis of our data along the lines of the complex dielectric constant
$\hat{\epsilon}(\omega)=\epsilon_1(\omega) + i \epsilon_2(\omega)$, 
or complex conductivity
$\hat{\sigma}=\hat{\epsilon}\,{\omega}/({4\pi i})$, respectively.

\subsection{Optical Conductivity}
\label{sec:optcond}
\begin{figure}
\includegraphics[width=80mm]{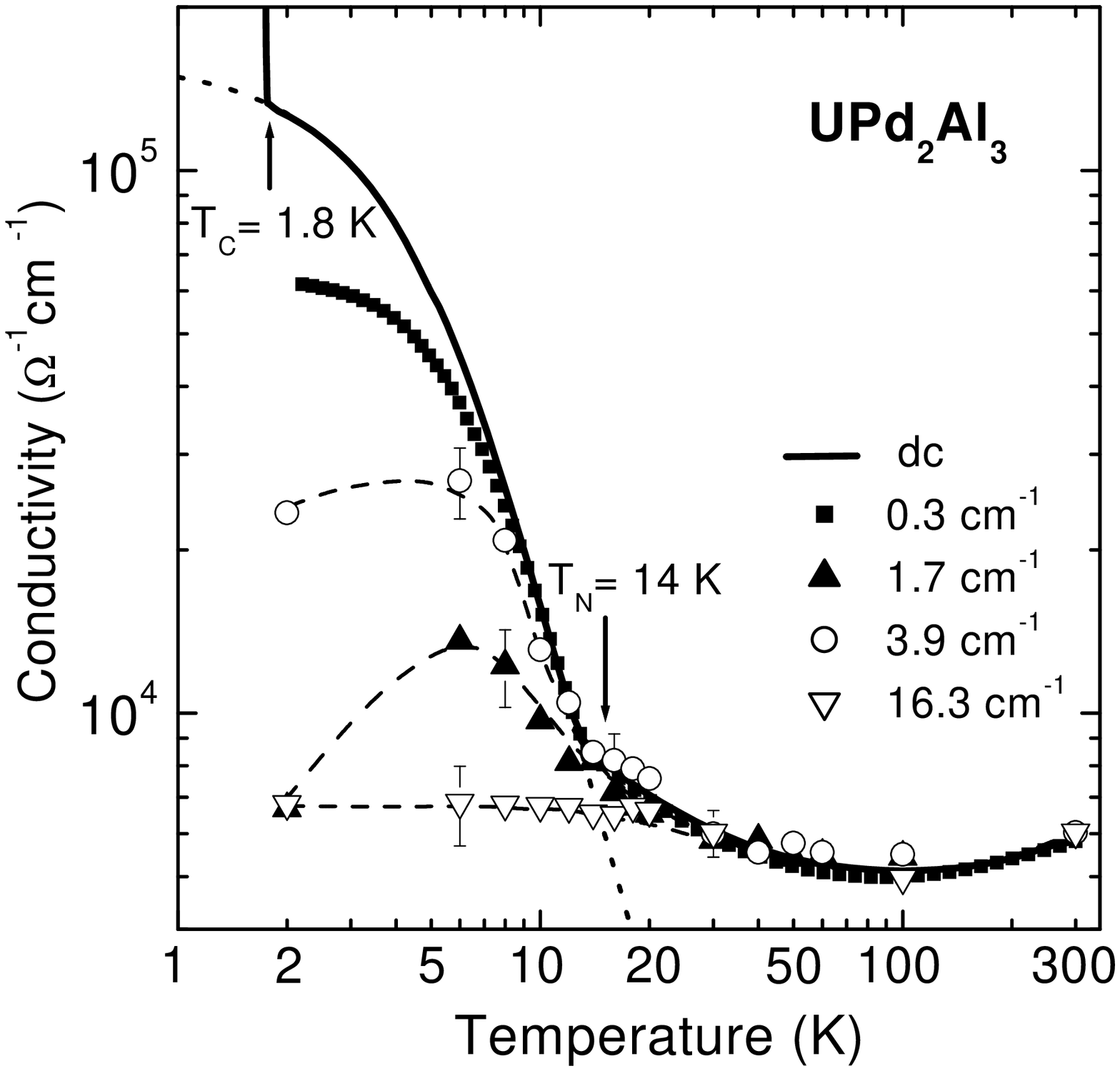}
\caption{\label{fig:figure1} Temperature dependence of the dc conductivity of the UPd$_2$Al$_3$ film together
with the ac conductivities obtained at different frequencies as indicated.  
The dotted line represents a fit of
the dc conductivity using Eq.~(\protect\ref{eq:magnon}) 
with the gap $E^{\prime}_g=1.9$~meV.
The dashed lines are guides
to the eye; note the non-monotonous frequency dependence at low temperatures.}
\end{figure}

\begin{figure}
\includegraphics[width=80mm]{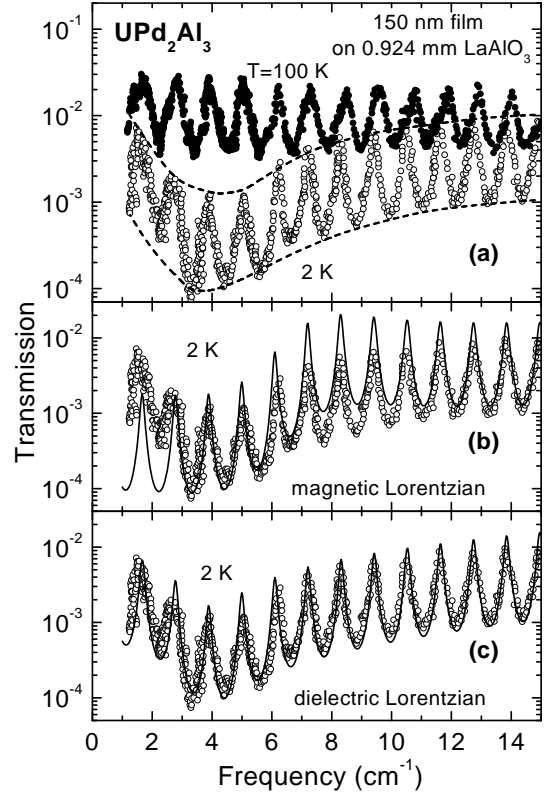}
\caption{\label{fig:figure2} Transmission spectra $T_F(\omega)$ of a 150~nm thick UPd$_2$Al$_3$ film on LaAlO$_3$
(thickness 0.924~mm) for temperatures $T=100$~K and 2~K (a); the dashed lines connecting minima and maxima (due to multi-reflection within the substrate) are
drawn to emphasize the overall frequency dependence of the transmission of the film. 
The middle (b) and the
bottom (c) panels present the fits of the transmission spectra for 
$T=2$~K using the models of `magnetic' and
`dielectric' oscillators (Lorentzians), respectively; 
the oscillator parameters are given in the text.}
\end{figure}

To discuss our findings on the temperature and frequency dependent transport 
in UPd$_2$Al$_3$,
let us first turn back to Fig.~\ref{fig:figure1}. Below $T_{N}$
 the temperature dependence of the dc
resistivity of the film can be described\cite{Dalichaouch92,Andersen80,Bakker93} using the expression for an
antiferromagnet with an energy gap $E^{\prime}_g$:
\begin{equation}
\rho(T)=\rho_0 + aT^2 + bT\left(1+\frac{2k_{\rm B}T}{E^{\prime}_g}\right)
\exp\left\{\frac{-E^{\prime}_g}{k_{\rm B}T}\right\} .
\label{eq:magnon}
\end{equation}
Here $\rho_0$ gives the residual resistivity and the second term
describes the electron-electron scattering (Fermi liquid). A fit
of the dc curve (dotted line in Fig.~\ref{fig:figure1}) by Eq.~(\ref{eq:magnon})
yields the gap value $E^{\prime}_g=1.9$~meV, which corresponds to the one
reported in Ref.~\onlinecite{Dalichaouch92,Huth93,Bakker93}.

From early tunneling measurements on UPd$_2$Al$_3$
a rather large spin-gap energy up to
12.4~meV was first suggested.\cite{Aarts94} Using thin
films of this compound, recent tunneling experiments in the
superconducting and normal state clearly demonstrate\cite{Huth00}
that there
remains a reduced DOS at zero-bias  up the temperature range of 7~K with
a gap value of 1.0~meV while the superconducting gap is at 0.235~meV.
The existence of a gap
in either the electronic DOS or in the magnon spectrum was opposed
by Caspary {\it et al.} \cite{Caspary93} and on grounds of
infrared measurements by Degiorgi {\it et al.} \cite{Degiorgi94}
In the light of the low-energy measurements presented in this
paper, the arguments will be reconsidered below.

\begin{figure}
\includegraphics[width=80mm]{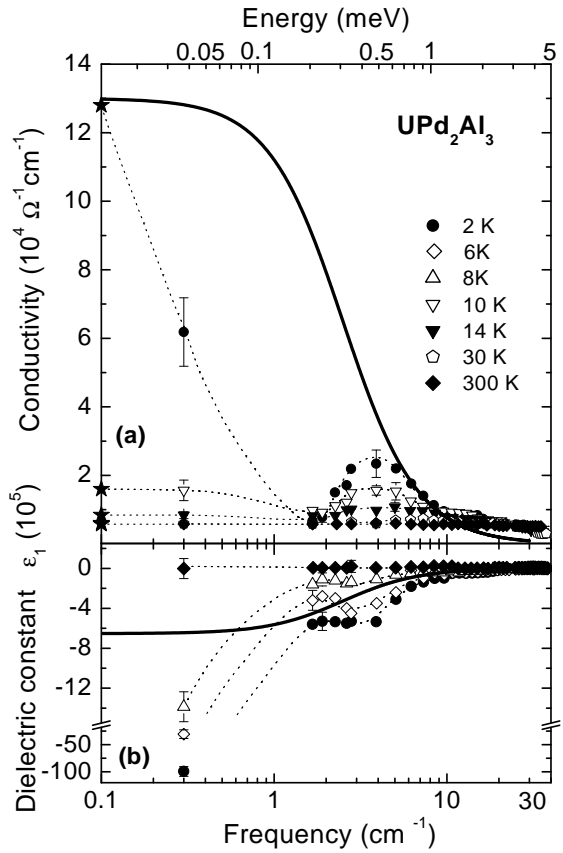}
\caption{\label{fig:figure3} The low-frequency optical conductivity 
(a) and the dielectric constant (b) of
UPd$_2$Al$_3$ as a function of frequency at several temperatures. 
The dc conductivity is shown by the stars on
the left axis of the upper panel.  The dashed lines are drawn to guide the eye. 
The solid lines correspond to the
renormalized Drude behavior [Eq.~(\protect\ref{eq:renDrude})] assuming  $\sigma_{dc}=1.25\cdot 10^5 
(\Omega{\rm cm})^{-1}$ and 
$\Gamma^*/(2\pi c) =2.5~{\rm cm}^{-1}$.}
\end{figure}

The real parts of the low-frequency optical conductivity $\sigma_1(\omega)$ 
and of the dielectric constant
$\epsilon_1(\omega)$ of UPd$_2$Al$_3$ are plotted in Fig.~\ref{fig:figure3} 
for some temperatures. Except for the
dc conductivity on the left axis of the upper frame and the results of the 
microwave impedance measurements at
0.3~cm$^{-1}$, all the data points plotted are obtained directly 
from the transmitted power and phase shift in
our transmission experiments. Within our accuracy both quantities are 
frequency independent for $T\geq 25$~K, but
show a strong dispersion for lower temperatures. In addition, the response 
of the Drude model [Eq.~(\ref{eq:renDrude})]
is shown (solid line) which matches the dc conductivity 
$\sigma_{dc}=1.28\cdot  10^{5}~(\Omega{\rm cm})^{-1}$ at
2~K and the far-infrared roll-off around 10~cm$^{-1}$; it is obvious 
that the observed optical response of  UPd$_2$Al$_3$
differs from that of a renormalized Drude metal. As a result, the simple 
Hagen-Rubens extrapolation --~in general
used \cite{Degiorgi99} to extrapolate the far-infrared reflectivity 
below 30~cm$^{-1}$~-- totally misses the
following two features in our millimeter and sub-millimeter range. 
First, the optical conductivity
$\sigma_1(\omega)$ clearly shows the development of a gap-like minimum 
below 3~cm$^{-1}$~at $T<20$~K; it
corresponds to a pronounced increase of the dielectric constant 
$\epsilon_1(\omega)$ which is a direct measure of
the gap. This feature gradually disappears with increasing temperature and 
is not seen above 30~K. Second, at
frequencies below approximately 1.5~cm$^{-1}$ the conductivity increases 
for $\omega\rightarrow 0$ towards
considerably higher dc values leading to a very narrow peak at zero frequency.

In a first approach we describe the narrow peak at $\omega=0$
by the Drude model
and use a phenomenological gap model \cite{DresselGruner,Yu96} for the higher
frequencies:
\begin{eqnarray}
\sigma_1(\omega)= \frac{\sigma_{dc}
(\Gamma^{*}_{D})^2}{\omega^2+(\Gamma^{*}_{D})^2} + (1-\Theta)
\frac{\Sigma_{g}}{\omega^2+\Gamma^{2}_{g}}\sqrt{\omega-\omega_{g}}
\label{eq:sigmagap}
\end{eqnarray}
where $\Theta=1$ for $\omega \leq \omega_{g}$ and $\Theta=0$ above the gap frequency $\omega_g$. Here $\Gamma^*_{D}$ is
the damping of the narrow $\omega = 0$ mode and $E_g=\hbar\omega_{g}$ 
corresponds to the  gap energy. The
 parameters $\Sigma_{g}$ and $\Gamma_{g}$ depend on the
concentration, the effective mass, and the scattering rate of the charge carriers 
above the gap. By performing a
Kramers-Kronig integration of Eq.~(\ref{eq:sigmagap}) 
we have also obtained an analytical expression for the dielectric
constant which was used in the fitting procedures. 
The assumed model (\ref{eq:sigmagap}) describes very well the
experimental results of $\sigma_1(\omega)$ and $\epsilon_1(\omega)$ 
for temperatures $T\leq20$~K; as an example,
the fit of the 2~K data is shown in Fig.~\ref{fig:figure4}. 
The temperature dependence of the gap energy
$\hbar\omega_{g}$ as obtained by this fit is displayed in the inset 
(solid squares corresponding to the left
axis). The gap value $E_{g}\approx 0.22$~meV is essentially temperature independent, which does not correspond to the well-known BCS-like 
temperature behavior obtained 
by mean-field theory; instead, the gap feature becomes more and
more pronounced as the temperature is lowered.

While the above analysis was solely based on direct measurements of the optical
conductivity and dielectric constant without applying the Kramers-Kronig analysis, 
we now want to combine all the absorption data available in the entire
range of frequency.
In Fig.~\ref{fig:figure3new}a the frequency dependence of the absorptivity
$A(\omega)=1-R(\omega)$ is plotted over a wide spectral range
and for different temperatures.
From the measured dc conductivity the absorptivity is calculated in the  Hagen-Rubens
limit assuming Eq.~(\ref{eq:hr}).  
The full symbols are the results of the microwave cavity measurements of the
surface resistance $R_S$, using Eq.~(\ref{eq:mw-abs-hr}). 
In the range from 1~cm$^{-1}$ to 40~cm$^{-1}$, 
the absorptivity is evaluated from our
optical experiments using the transmission coefficient and the phase shift (open circles; 
only 2~K data are
plotted for clarity reasons). Above 30~cm$^{-1}$ up to $10^6~{\rm cm}^{-1}$ 
we also utilized the data from bulk
reflectivity measurements  (solid lines) published 
previously.\cite{Degiorgi97,Degiorgi94} For even higher
frequencies we extrapolated by $R(\omega)\propto \omega^{- 4}$. 
The lines represent the input we used for
performing the Kramers-Kronig analysis in order to calculate the optical conductivity. 
It was obtained by
simultaneously fitting the absorptivity results and the directly measured conductivity. Obviously, the
low-temperature data exhibit significant deviations from the $A(\omega)\propto\omega^{1/2}$ Hagen-Rubens behavior
for frequency above 0.1~cm$^{-1}$.

The lower panel of  Fig.~\ref{fig:figure3new} shows the corresponding optical 
conductivity $\sigma_1(\omega)$ as
obtained by a Kramers-Kronig analysis of the absorptivity just described; 
in addition we again display the
directly measured conductivity data which were also taken into account for the 
absorptivity fit, as mentioned
above. The agreement is very good considering the error bars for each 
measurement technique. The main features
can be summarized as follows: 
(i)~At high temperatures $T>50$~K, we observe 
a broad conductivity spectrum as
expected for a metal according to Eq.~(\ref{eq:metalDrude}). Following Ref.~\onlinecite{Degiorgi94},  the plasma
frequency from the evaluation of the entire spectral weight of all free charge 
carriers is $\omega_p/(2\pi c)=
4.4\cdot 10^4~{\rm cm}^{-1}$. (ii) As the temperature decreases below $T^*$, 
a renormalized Drude peak
[Eq.~(\ref{eq:renDrude})] develops due to the gradual enhancement 
of the effective mass $m^*$. The renormalized
plasma frequency $\omega_p^*/(2\pi c) \approx$ 4350 cm$^{-1}$, 
which will be obtained by the procedure discussed
in Subsection~\ref{sec:spectralweight}, corresponds to $m^*\approx 100 m$. 
This behavior is typical for heavy
fermions \cite{Degiorgi99} and was discussed in detail by Degiorgi, Dressel and 
coworkers
\cite{Degiorgi97,Degiorgi94,Dressel98,Dressel00} for the case of UPd$_2$Al$_3$. 
At the same time, a gap-like
feature develops around 100~cm$^{-1}$ as expected from the hybridization of the 
localized $5f$  electrons and the
conduction electrons; this value is in good agreement with the scaling relation (\ref{eq:scaling}). 
(iii)~Lowering the temperature further, magnetic ordering sets in at 
$T_N=14$~K and leads to distinct signatures
in the optical spectrum. 
We observe a well pronounced pseudogap below 2~cm$^{-1}$ 
which we assign to magnetic
correlations between the localized and delocalized charge carriers. 
At even lower frequencies, an extremely narrow mode 
centered zero-frequency remains which is eventually responsible 
for superconductivity below $T_c=2$~K. The analysis and
understanding of this low-temperature ($T_c<T\leq T_N$) and low-frequency ($\nu<50$~cm$^{-1}$) behavior is the
main point of this paper.

For the analysis of our low-frequency data we introduce a complex frequency 
dependent scattering rate
$\hat{\Gamma}(\omega)=\Gamma_1(\omega)+i\Gamma_2(\omega)$ into the 
standard Drude form of
Eq.~(\ref{eq:renDrude}). If we define the dimensionless quantity 
$\lambda(\omega)=-\Gamma_2(\omega)/\omega$, the
complex conductivity can be written as
\begin{equation}
\hat{\sigma}(\omega)=\frac{(\omega_p^{\prime})^2}{4\pi}
\frac{1}{\Gamma_1(\omega)-i\omega(m^*(\omega)/m )}
\label{eq:gen-drude2}
\end{equation}
where $m^*/m =1+\lambda(\omega)$ is the frequency dependent enhanced mass.
$\omega_p^{\prime}/(2\pi c) = 9.5 \cdot 10^3~{\rm cm}^{-1}$ corresponds to the 
fraction of electrons
which participate in the many body state which develops below $T^*$ as discussed 
in the following subsection.
By rearranging Eq.~(\ref{eq:gen-drude2}) we can write expressions for
$\Gamma_1(\omega)$ and $m^*(\omega)$ in terms of $\sigma_1(\omega)$ and
$\sigma_2(\omega)$ as follows:
\begin{equation}
\Gamma_1(\omega)=\frac{(\omega_p^{\prime})^2}{4\pi}
\frac{\sigma_1(\omega)}{|\hat{\sigma}(\omega)|^2}
\label{eq:gam-w}
\end{equation}
\begin{equation}
\frac{m^*(\omega)}{m}=\frac{(\omega_p^{\prime})^2}{4\pi}
\frac{\sigma_2(\omega)}{|\hat{\sigma}(\omega)|^2}\frac{1}{\omega} .
\label{eq:mstar-w}
\end{equation}
Due to causality, \cite{allen_drude} $\Gamma_1(\omega)$ and $m^*(\omega)$ are 
related through the Kramers-Kronig
integrals. \cite{DresselGruner} Such analysis  allows us to look for interactions 
which would lead to energy
dependent renormalization effects as the frequency dependent scattering rate and 
effective mass are related to the
real and imaginary parts of the frequency dependent self-energy of the electrons.\cite{Abrikosov65} Such
analysis has been used before in studying the response of HF compounds \cite{Dordevic01,Sulewski88,Awasthi93} and
high temperature superconductors. \cite{ruvalds_nested_fl} In the lower frames of
Fig.~\ref{fig:figure7new} the frequency dependence
of the scattering rate $\Gamma_1(\omega)$ and of the effective mass $m^*(\omega)$ 
are displayed for different
temperatures; for comparison we plot $\sigma_1(\omega)$ and $\epsilon_1(\omega)$ 
in Fig.~\ref{fig:figure7new}a
and b.  As expected for a Drude metal, at $T>T^*$ the spectra of 
$\Gamma_1(\omega)$ and $m^*(\omega)$ are nearly
frequency independent. As the temperature is lowered we observe a peak in the 
energy dependent scattering rate at
the hybridization gap, as can be nicely seen in the $T=30$~K data of Fig.~\ref{fig:figure7new}c. The existence of
HF quasi-particles is confirmed by a low-frequency plasmon characterized by the 
zero-crossing in the spectra of
the dielectric constant $\epsilon_1(\omega)$ shown in Fig.~\ref{fig:figure7new}b. 
The decrease of
$\Gamma_1(\omega)$ to lower frequencies corresponds to an increase of 
$m^*(\omega)$. At these intermediate
temperatures, $T_N<T<T^*$, the effective mass $m^*/m$ already reaches 
a value of 35.

The question remains how the optical response at very low frequencies can be 
described, i.e.\ for excitations of the zero-frequency mode;
whether it is simply a renormalized Drude behavior.
The Landau Fermi-liquid theory \cite{pines_nozieres_book} predicts that
the scattering rate due to electron--electron interactions in three
dimensions should be quadratic both in temperature and
frequency. \cite{ruvalds_nested_fl,pines_nozieres_book,Ashcroft76,gurzhi_scattering_fl}
In order to examine the shape of the zero-frequency peak observed in the
data, we have adopted the following phenomenological forms of 
$\Gamma_1(\omega)$ and
$m^*(\omega)$, used by Sulewski~{\em et al.} in their study of the Fermi liquid
behavior of the heavy fermion compound UPt$_3$: \cite{Sulewski88}
\begin{equation}
\Gamma_1(\omega)= \Gamma_0 + \frac{\lambda_0 \alpha \omega^2}
{1+ \alpha^2\omega^2} \label{eq:gamma}
\end{equation}
and
\begin{equation}
\frac{m^*(\omega)}{m} = 1+ \frac{\lambda_0}{1+\alpha^2\omega^2},
\end{equation}
where $\Gamma_0$ and $\lambda_0$ are the zero-frequency scattering rate and mass enhancement,
respectively. $1/\alpha$ is a characteristic frequency (energy) of the process. These expressions obey the
Kramers-Kronig relation and have the proper Fermi-liquid frequency dependence. 
The comparison of the low-frequency behavior
of $\Gamma_1(\omega)$ at $T=30$~K to the fit by Eq.~(\ref{eq:gamma}) 
(inset of Fig.~\ref{fig:figure7new}) confirms
the main features: as the frequency increases, $\Gamma_1(\omega)$ grows and 
then saturates around 100~cm$^{-1}$.
The present set of data, however, does not allow to conclude whether 
UPd$_2$Al$_3$ can be satisfactorily
described within a
Fermi-liquid formalism or not. These questions remain the subject of further investigations
at lower frequencies.

The scattering rate drops more than one order of magnitude below 15~K. This can be explained by the freezing-out
of spin-flip scattering below $T_N$. As was recently pointed out \cite{Marsiglio01,Basov02} in the context of
high temperature superconductors, the energy dependent scattering rate is related 
to the electronic DOS. In fact,
a sum-rule was suggested similar to Eq.~(\ref{eq:metalsumrule}). In the case of UPd$_2$Al$_3$ we can distinguish
two gap structures with an enhanced DOS at the edges, as it is known from superconductors and low-dimensional
semiconductors. Again, it would be of great interest to determine 
$\Gamma_1(\omega)$ of this narrow
zero-frequency mode which remains in the AF state. Due to
the influence of magnetic correlations it might be well distinct from 
the behavior at $T>T_N$ plotted in the inset of Fig.~\ref{fig:figure7new}. 
The related experiments are currently in progress.

Decreasing the temperature below $T_N$ results in a second peak of 
$\Gamma_1(\omega)$ and a strong increase of
the effective mass around 1~cm$^{-1}$. Already seen at $T=15$~K, 
this effect becomes stronger as the temperature
is lowered to 2~K. The effective mass $m^*/m$ levels off below this frequency and 
nicely matches the values
obtained by thermodynamic methods \cite{Geibel91,Dalichaouch92} 
which are indicated by the point at $\omega=0$ in
Fig.~\ref{fig:figure7new}d. Note, that due to the few data points, 
the features below 1~cm$^{-1}$ are not
significant; this range demands further studies. 
The important finding of our study is, that  the magnetic ordering significantly 
changes the energy dependence
$m^*(\omega)$: the strong increase of the effective mass does not occur at the 
hybridization gap at 10~meV, but
mainly below the energy range which corresponds to the correlation gap at 0.2~meV. 
This infers that the magnetic
correlations are of superior importance for the mass enhancement.

\subsection{Spectral Weight}
\label{sec:spectralweight}
According to Eq.~(\ref{eq:metalsumrule}),
the area under the conductivity spectrum is related to the
 density and the mass of the charge carriers. 
Well above the coherence temperature $T^*$, optical spectra of 
UPd$_2$Al$_3$
 simply follow the Drude behavior of a normal metal with an unrenormalized  
plasma frequency of $\omega_p/(2\pi c)= 4.4\cdot 10^4~{\rm cm}^{-1}$ as 
evaluated previously.\cite{Degiorgi94} Assuming $m= m$, the 
free electron mass, we  estimate
a charge carrier density $n= 1.1\cdot 10^{22}~{\rm cm}^{-3}$. From
experimental \cite{Inada94} and calculated \cite{Sandratskii94} de
Haas--van Alphen spectra we know that only approximately 35~\%\ of the
electrons contribute to the HF state; qualitatively we observe this fact also in our 
optical spectra.
The plasma frequency which corresponds to those electrons 
(concentration $n^{\prime}$) which are affected by the formation
 of the HF state \cite{remark1} is 
$\omega_p^{\prime}/(2\pi c)=9.5\cdot 10^3$~cm$^{-1}$.
 Our analysis further infers that these carriers have a band mass $m^{\prime}=7m$. 
This also means,
that a large fraction of the carriers is unaffected by the HF ground state and
will not be considered in our further analysis.

\begin{figure}
\includegraphics[width=80mm]{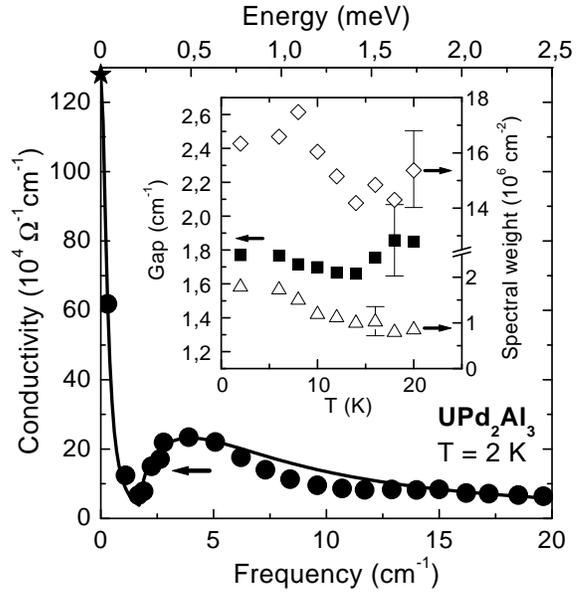}
\caption{\label{fig:figure4} Optical conductivity $\sigma_1(\omega)$ of UPd$_2$Al$_3$ 
at $T=2$~K. The  line corresponds to
the fit by Eq.~(\ref{eq:sigmagap}). \label{fig:figure5} 
In the inset the temperature dependences of the 
gap frequency (full squares, left axis) obtained from the 
fit is displayed. 
Also shown is the spectral
weight of the $\omega =0$ resonance (open triangle, right axis) and 
of all the excitations above and below the
gap (open diamonds, right axis). For details see text.}
\end{figure}

\begin{figure}
\includegraphics[width=80mm]{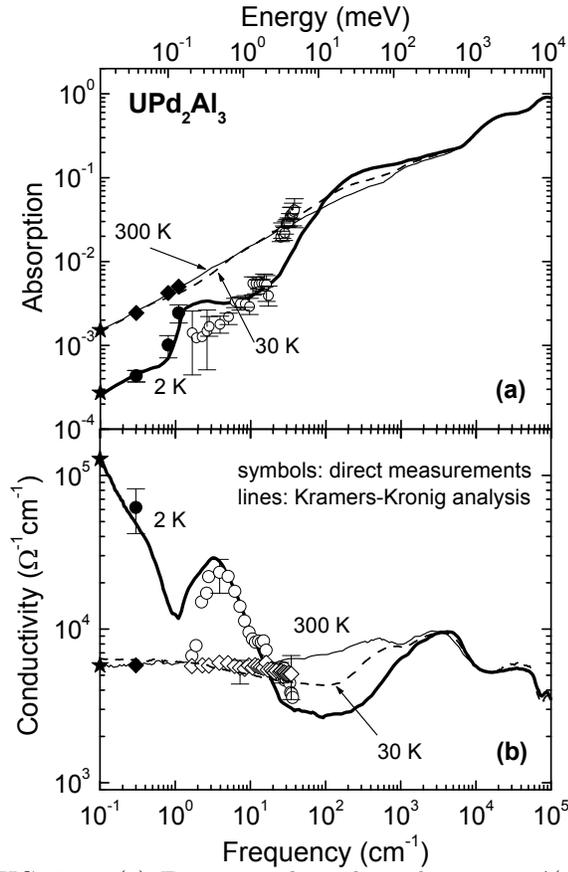}
\caption{\label{fig:figure3new} (a) Frequency dependent absorption $A(\omega)$ 
of UPd$_2$Al$_3$ at different temperatures  shown over a wide frequency range.
The solid stars
on the left axis represent the dc values in a Hagen-Rubens behavior 
(\protect\ref{eq:hr}); 
the full symbols in
the microwave range are obtained by cavity perturbation technique; 
the open symbols present absorption evaluated
from the transmission and phase measurements by the Mach-Zehnder 
interferometer (only 2~K data are plotted for
clarity reasons). The lines are obtained by combining the various optical 
investigations (transmission through
films and reflection of bulk samples) and simultaneously matching the directly 
measured conductivity and
dielectric constant. (b) Optical conductivity of UPd$_2$Al$_3$ as evaluated by a 
Kramers-Kronig analysis of the
above absorptivity data. The stars on the left axis indicate the dc conductivity; 
the full points were obtained by the 10~GHz cavity method. 
The open  symbols correspond to the direct determination of 
the optical conductivity using the
transmission and the phase shift obtained by the Mach-Zehnder interferometer.}
\end{figure}

\begin{figure}
\includegraphics[width=80mm]{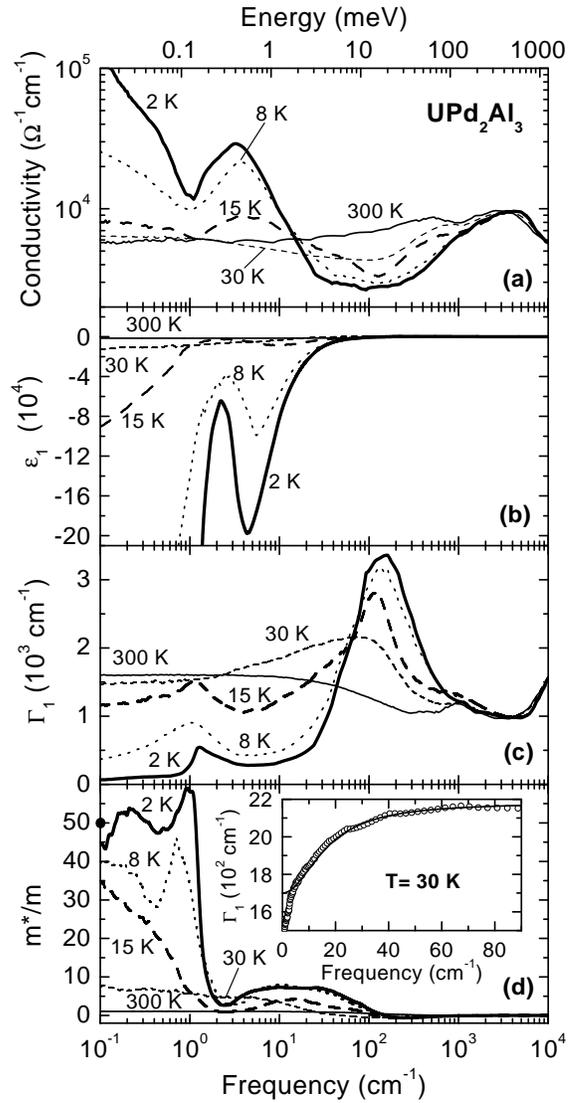}
\caption{\label{fig:figure7new}Frequency dependence of (a) the optical conductivity $\sigma_1(\omega)$, (b) the
dielectric constant $\epsilon_1(\omega)$, 
(c) the scattering rate $\Gamma_1(\omega)$, 
and (d) the effective mass
$m^*(\omega)/m$ of UPd$_2$Al$_3$ for different temperatures. 
The point on the left axis of panel (d) corresponds to
the effective mass determined by thermodynamic measurements 
(data from Refs.~\protect\onlinecite{Geibel91,Dalichaouch92}), 
and the inset shows the comparison of $\Gamma_1(\omega)$ at
$T=30$~K with the behavior (\ref{eq:gamma}) expected for Landau's 
theory of a Fermi-liquid.}
\end{figure}

Typical for a HF compound, below 100~cm$^{-1}$ a narrow conductivity mode 
develops due to electronic correlations
at lower temperatures. For $T>20$~K but below $T^*$, the spectral weight of 
this contribution does not change
significantly; the renormalized plasma frequency is $\omega_p^*/(2\pi c) \approx 
4350$~cm$^{-1}$. A similar
result of a temperature independent $\omega_p^*$ was obtained from our fit of the 
directly measured conductivity
as displayed in the inset  of the Fig.~\ref{fig:figure5} by the open diamonds.

Assuming that
the total number of charge carriers $n^{\prime}$ remains unchanged,
sum-rule arguments give $\omega_p^{\prime}/\omega^*_p = 
\sqrt{m^*/m^{\prime}}$.
Thus we obtain $m^*/m \approx 30$
which is in good agreement with the value obtained by thermodynamic
methods of specific heat and susceptibility ($m^*/m =41-66$) \cite{Geibel91,Dalichaouch92} and
is somewhat smaller than the estimate from
infrared measurements ($m^*/m=85$).\cite{Degiorgi94} Analyzing our data by the generalized Drude model
(\ref{eq:gen-drude2}), we have obtained the energy dependence of
the effective mass as plotted in Fig.~\ref{fig:figure7new}d.
For $T=2$~K we find an enhancement of $m^*(\omega)/m=45$, in excellent
agreement with the above estimates of the spectral weight.

As the temperature decreases below $T_N$, a  pseudogap develops in the optical 
spectra around 1.8~cm$^{-1}$
(i.e.\ 0.2~meV) with important implications on the spectral weight. 
At low temperatures the optical weight
consists of two contributions, the narrow $\omega$=0 centered resonance and 
the charge carriers excited across the
gap of energy $\hbar\omega_g$
\begin{equation}
\int_0^{\omega_c} \sigma_1(\omega)\,{\rm d}\omega =
\int_0^{\omega_g} \sigma_1(\omega)\,{\rm d}\omega +
\int_{\omega_g}^{\omega_c} \sigma_1(\omega)\,{\rm d}\omega .
\label{eq:OW}
\end{equation}
Here we integrated up to a cut-off frequency $\omega_c/(2\pi c)\approx 
100$~cm$^{-1}$ which lies above the
features associated to correlation effects but well below the interband transitions. 
The zero-frequency response
contains only 10\%\ of the overall spectral weight, the major contribution comes 
from the excitations across the
gap. This agrees well with the above discussion of the peak in the DOS 
right above the gap. The temperature
dependences of both components as well as their sum are presented in the inset  
of Fig.~\ref{fig:figure5} (right
axis). The contribution of the $\omega=0$ response increases as the temperature 
decreases (open triangles in inset of
Fig.~\ref{fig:figure5}); within our accuracy, this however, does not lead to a 
significant change of the overall
spectral weight with temperature (open diamonds). The observed behavior can be 
taken as an indication that these
two modes correspond to charge carriers localized at different parts of the Fermi 
surface. The coexistence of two
different electronic subsystems in UPd$_2$Al$_3$ has also been suggested in Ref.\onlinecite{Caspary93}.

The most interesting question is the relation between the AF
ordering and the pseudogap origin. The fact that we also see the
gap-like feature slightly above $T_{\rm N}$ up to 
20~K or more does not rule out its connection
to the magnetic ordering since an incommensurate
phase was also observed up to $T \approx 20$~K by neutron diffraction
experiments. \cite{Krimmel92}

\section{Discussion}
Since the development of the HF state in UPd$_2$Al$_3$ 
and the corresponding hybridization gap, 
which we observe around 100~cm$^{-1}$, 
was already discussed  in Ref.~\onlinecite{Degiorgi99,Degiorgi94} extensively, 
we concentrate now
on the newly discovered feature at lower energies.
Four possible explanations for the origin of the correlation gap at 0.2~meV
 will be considered:
(i) the pseudogap in the optical response may be related to spin-wave excitations, 
(ii) the formation of a SDW
ground state may lead to the opening of an energy gap, (iii) due to interaction the hybridization gap is shifted
to low energies, and (iv) magnetic correlations in the AF ground state influence the electronic DOS spectra.

(i) From torque magnetization measurements S\"ullow {\it et al.} \cite{Sullow96} found evidence of a 
gap in the spin-wave
spectrum of the order of $E_g\approx 0.4$~meV, which is about the value of 
the pseudogap we see in the optical
spectra. For $T>T_c=2$~K inelastic neutron scattering experiments found a 
Lorentzian-shaped line around 1.5~meV
which is ascribed to magnetic excitations. \cite{Bernhoeft98} 
Another  gap observed at 0.4~meV was associated
with superconductivity. All of these modes exhibit a strong temperature and 
{\bf q}-dependence. \cite{Bernhoeft98}
Besides the resistivity $\rho(T)$ discussed above, the fall of $C(T)$ was also 
interpreted as the opening of a
gap in the magnetic excitation spectrum. \cite{Geibel91,Caspary93} 
Typically magnetic excitations, however, lead
to a much smaller response in the optical properties. As it was shown above 
in
Sec.~\ref{sec:transmissionspectrum}, we were not able to obtain a reasonable fit 
of the observed features by a
magnetic absorption process (cf.\ Fig.~\ref{fig:figure2}). Moreover, in our 
preliminary experiments in magnetic
field \cite{Kasper01} we have not seen any clear evidence in favor of the magnetic 
nature of the observed gap
feature. We also think that the dielectric constant $\epsilon_1(\omega)$ would show 
a stronger influence of the
magnetic field if the features were related to an AF resonance; in particular the 
frequency of the excitation would
shift.

(ii) Below $T_N$ a commensurate ordering of the rather large
magnetic moments (0.85~$\mu_B$) occurs in UPd$_2$Al$_3$;\cite{Krimmel92}
also in the HF antiferromagnet URu$_2$Si$_2$ a commensurate structure develops 
but the ordered moment only
amounts to 0.02~$\mu_B$. Soon after the discovery of UPd$_2$Al$_3$ it was 
debated whether also in this compound
the antiferromagnetic
ordering is due to the SDW ground state like in URu$_2$Si$_2$. In contrast to URu$_2$Si$_2$, the resistivity of
UPd$_2$Al$_3$ does not increase right below $T_N$, but the low temperature 
decrease can be well described by an
activated  behavior [Eq.~(\ref{eq:magnon})]. We do not think that the so-obtained 
gap value $E_g$=1.9 meV
indicates the formation of a SDW state \cite{Dalichaouch92,Bakker93}; 
it is much larger than the value obtained
from our optical results (Fig.~\ref{fig:figure4}). The pseudogap feature we see in UPd$_2$Al$_3$ is not as
clearly developed as the energy gap in URu$_2$Si$_2$ or  UCu$_5$ \cite{Bonn88,Degiorgi94} and about a factor of
five below the frequency one would expect from mean-field theory 
$E_g = 3.5k_BT_N$. The opening of a SDW gap can
be seen in the temperature dependent NMR relaxation rate of URu$_2$Si$_2$ \cite{Kohara86} yielding a reduction of
the electronic density of states by a factor of three. A very similar behavior has been detected by $^{27}$Al-NQR
experiments \cite{Tou95} in UPd$_2$Al$_3$. Indications for a SDW behavior in isostructural UNi$_2$Al$_3$ have
been reported by Y. Dalichaouch {\it et al.} \cite{Dalichaouch92} and by 
Uemura {\it et al.} \cite {Uemura93}
Recent $^{105}$Pd-NMR and NQR experiments, however,  speak against 
the SDW model for the UPd$_2$Al$_3$; the
observed divergence in $1/T_1$ at $T_N$ can be explained by a localized moment 
picture of uranium.\cite{Matsuda97}

(iii)  Another  possible mechanism of the pseudogap formation at these low energies  
is the change of the electronic
DOS due to the coherence of screening conduction electrons in the Kondo lattice 
(so-called hybridization gap
\cite{Grewe91,Wachter93}). 
The coherence temperature $T^*$ is usually of the same
order of magnitude as  the single particle Kondo temperature $T_K$, 
\cite{Gruner78}
 below which local moments are
screened and a local Fermi-liquid picture applies. According
to different experiments, \cite{Geibel91,Geibel912,Caspary93}
 the
coherence temperature of UPd$_2$Al$_3$ lies in the range from 20 K to 60 K, i.e.\ 
the hybridization gap should be
approximately 1.9 to 5~meV. This is in excellent agreement with the 100~cm$^{-1}$ 
gap (124~meV) we observe and hence assign
to the hybridization gap following Eq.~(\ref{eq:scaling}). However, there might be 
scenarios which also relate the
low-frequency gap at 0.2~meV to the hybridization:
\\
(a) The strong anisotropy in UPd$_2$Al$_3$ may lead to an anisotropic 
hybridization gap, and  optical measurements
see  an ``effective'' value, which can be lower than the single-particle Kondo coupling energy. We want to recall a
similar discussion for high-temperature superconductors where a complicated gap symmetry is made responsible for
`states in the gap' and a distinct optical response. \cite{Graf95,Quinlan96,Schuerrer98} However, such a situation
is only realized for $f^1$ or $f^3$ configurations of the U-ions, due to the absence of 
the Kondo screening for
$f^2$ state. On the other side many experiments indicate U$^{4+}$ oxidation state 
($f^2$-configuration) in
UPd$_2$Al$_3$, \cite{Steglich91} but the electronic configuration of the U is not determined
unambiguously.\cite{Paolasini93}
\\
(b) By means of analytical and numerical cal\-cul\-a\-tions \cite{Burdin2000,Tahvildar98} the ratio $T_K/T^*$
should be less than unity in case of a low density of screening electron  
(exhausting  regime). It leads to a 
decrease of the coherence temperature $T^*$, which is a measure of the hybridization 
gap. In this case our
Kondo-system would have single-particle excitations with an energy of the order $k_BT_K\approx$1.9 meV 
[it corresponds to the energy of the mode (1.5~meV), observed in the inelastic 
neutron scattering experiments
\cite{Bernhoeft98}] 
and a collective resonance in the heavy particle assembly with about 
ten times lower energy. In
favor of this hypothesis speaks the high magnetic moment of U (low concentration of 
the screening electrons) and
also the fact that we observe the pseudogap up to $T\approx 20$~K: 
as the numerical calculations in the
periodic Anderson model indicate, the Kondo resonance in the exhausting regime shows 
a much weaker temperature
dependence and can be seen up to temperature $T \approx 10\,T^*$, in full agreement 
with our experimental
observation.

(iv) The itinerant antiferromagnetism of a SDW is at
odds with the formation of local moment magnetism deduced from
susceptibility \cite{Geibel91} and neutron scattering.\cite{Krimmel92}
On the other hand, using self-consistent density
functional calculations in the local approximation the magnetic
structure and the size of the ordered moment of UPd$_2$Al$_3$ have
been well described within a purely itinerant electron picture.\cite{Sandratskii94}
 It was suggested \cite{Caspary93} that two
different  electronic subsystems coexist in UPd$_2$Al$_3$. One of
them is a rather localized uranium $5f$ state responsible for the
magnetic properties, the other is delocalized and determines
the heavy fermion and superconducting properties. From the London
penetration depth \cite{Caspary93} $\lambda_L(0)$=450 nm we
calculate \cite{Tinkham96} the plasma frequency of the superconducting
carriers $(2 \pi \lambda_L(0))^{-1}=3540$~cm$^{-1}$ and find a
good agreement with $\omega_p^*/(2\pi c)=4350$~cm$^{-1}$
obtained from the spectral weight of our low-frequency
conductivity spectrum just above $T_c$ (cf.\ inset of Fig.\ref{fig:figure5}).
This has the following implications:\\
(a) In order to recover
the spectral weight $\rho^s$ of the $\delta$-peak in the superconducting state
(as determined from the penetration depth) according to
the Tinkham-Ferrell sum rule, \cite{DresselGruner,Tinkham96}
we have to integrate the normal state conductivity
$\sigma_1^n(\omega)$ up to a cut-off frequency $\omega_c^s/(2\pi c)
=100~{\rm cm}^{-1}$ (assuming that the conductivity decreases to zero up to this frequency) which is well above the frequency $2\Delta/(hc) =
4~{\rm cm}^{-1}$ at which the superconducting gap was
observed by tunneling spectroscopy.\cite{Jourdan99}
In conventional superconductors a change in the optical properties upon
entering the superconducting state
can be observed only up to approximately three times the gap frequency.
A similar discrepancy of spectral weight was found in 
high-temperature superconductors  \cite{Basov99}
and gave argument to a change of the kinetic energy $\Delta K$ associated with 
the superconducting transition: \cite{Basov99,Hirsch92}
\begin{equation}
\rho^s=\int_{0+}^{\omega_c^s}\left[\sigma_1^n(\omega) - 
\sigma_1^s(\omega)\right]{\rm d}\omega +\Delta K.
\end{equation}
Our results indicate that also for UPd$_2$Al$_3$ there 
is a loss of kinetic energy in the superconducting state.
Since we were not yet able to probe the superconducting state and determine the superconducting gap by optical means,\cite{Basov01}  we do not want to  
further speculate on the violation of the Tinkham-Ferrell sum rule and 
possible implications.\\
(b) All electrons seen in our low-energy
spectra are in the HF ground state and eventually undergo the
superconducting transition below $T_c$. We can definitely rule
out an assignment of the gap to the localized carriers of the AF
ordered states with the delocalized carriers contributing only to
the narrow feature around $\omega=0$ because with a plasma frequency
of 1500~cm$^{-1}$ at low temperatures, this feature contains
only 18 \%\
of the carriers which become superconducting.
The small spectral weight below the correlation gap
implies that
the excitations above the gap stem from the delocalized states
and that the pseudogap observed in our conductivity spectra is
either inherent to the heavy-quasiparticle state or it is related
to exchange correlations of the second subsystem.

Recently Sato {\em et al.} \cite{Sato01} argued that the magnetic excitations
seen by neutron scattering produce effective interactions between itinerant
electrons, and therefore are responsible for superconductivity.
Similar  conclusions where drawn from tunneling measurements on 
UPd$_2$Al$_3$ films.\cite{Jourdan99} Our results now indicate that 
already in the normal state
electronic and magnetic excitations interact in the energy range
which is relevant for superconductivity in UPd$_2$Al$_3$. From the frequency
dependence of  the effective mass $m^*(\omega)$ 
we can conclude that
also the mass enhancement in the metallic state
is strongly influenced by the magnetic correlations.
Thus as magnetic excitons are supposed to be responsible for
the superconductivity \cite{Jourdan99,Sato01}, we  suggest that the
localized magnetic excitations also influence the properties of the normal state
which were commonly assigned to the heavy-fermion ground state.
We propose that  the magnetic order in these compounds is
the pre-requisite to the formation of the  heavy quasiparticles
and eventually of superconductivity.

It is expected that the {\em temperature dependence} of the quasiparticle formation, 
as observed in the enhancement of the effective mass, goes hand in hand 
with the {\em energy dependence} of the correlation effects, as probed by the f
requency dependence of $m^*$; however, only very few experiments 
on heavy fermion systems have been performed in this
 regard.\cite{Sulewski88,Awasthi93} 
From Fig.~\ref{fig:figure7new}d we
see a moderate enhancement of $m^*(\omega,T=0)$ for frequencies below 
the hybridization gap
and a larger one below the gap due to magnetic correlations. 
Accordingly, $m^*(\omega=0,T)$ increases
only slightly below $T^*$ and shows a large enhancement below $T_N$. 
The latter behavior should also
be observed in thermodynamic measurements of $m^*$.

\section{Outlook}
If our  explanation the optical properties of 
UPd$_2$Al$_3$ is valid, it may also apply to other heavy fermion 
compounds, in particular to UPt$_3$ which exhibits a similar behavior 
in many regards \cite{Ott87,Stewart84,Fisk88}.  For UPt$_3$ the effective 
mass of the quasiparticles is larger ($m^*/m\approx 200$) and the relevant 
energy scales are lower. The coherence temperature $T^*\approx 30$~K, 
fluctuating short-range magnetic order  
occurs at $T_N=5$~K,\cite{Kjems88}
 and superconductivity sets in at $T_c=0.5$~K. \cite{Stewart84} 
Although the  magnetic moment ($0.02\mu_B$) is much smaller, 
recent 
bandstructure calculations \cite{Zwicknagl01} infer the existence of localized as well 
as delocalized $5f$-electrons in UPt$_3$ very much 
similar to UPd$_2$Al$_3$. It was suggested that the observed enhancement of 
the quasiparticle mass results from the local exchange interaction of two 
localized $5f$-electrons with the remaining delocalized ones.

\begin{figure}
\includegraphics[width=80mm]{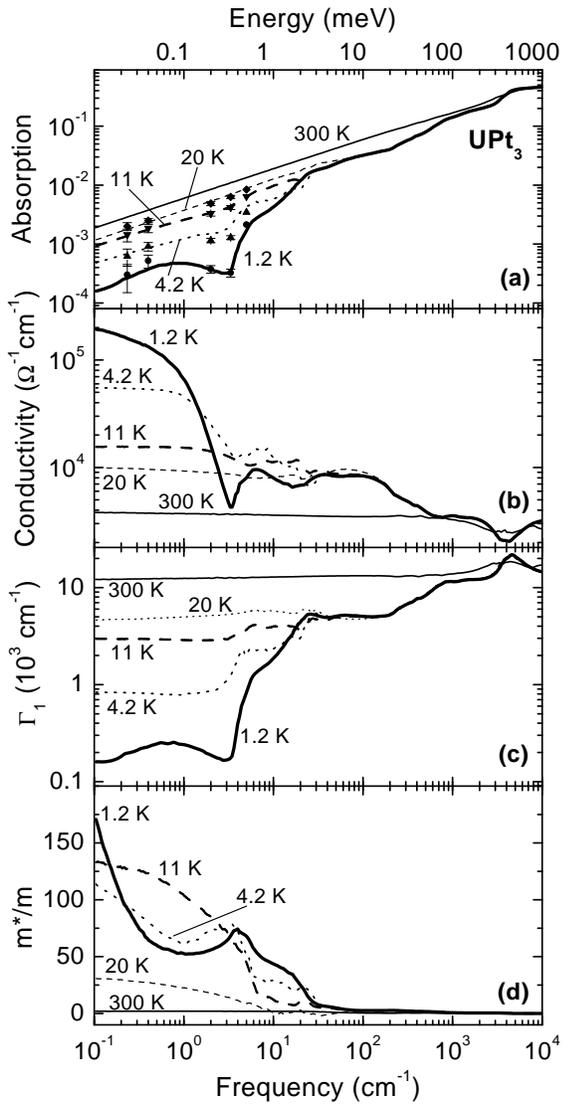}
\caption{\label{fig:UPt3}Frequency dependence of (a) the absorption $A(\omega)$, 
(b) the optical conductivity $\sigma_1(\omega)$,  (c) the scattering rate $\Gamma_1(\omega)$, and (d) the effective mass
$m^*(\omega)$ of UPt$_3$ for different temperatures. The absorption data are 
calculated form the experimental results 
of Refs.~\protect\onlinecite{Donovan97,Sulewski88,Marabelli86}.}
\end{figure}

Millimeter wave experiments on UPt$_3$ crystals
\cite{Donovan97,Sulewski88,Marabelli86} which show a peak in the conductivity 
at 6~cm$^{-1}$ for temperatures below 5~K indicate a similar scenario. 
We reanalyzed the  optical properties of
UPt$_3$ shown in Fig.~\ref{fig:UPt3}a in the same way as described above (Sec.~\ref{sec:optcond})
and obtain similar features. \cite{Dressel02}
When the coherent ground state builds up ($T<T^*$), the optical conductivity
increases for frequencies below 30~cm$^{-1}$ (Fig.~\ref{fig:UPt3}b).
As the temperature drops below $T_N=5$~K, magnetic ordering occurs  and an 
energy gap progressively develops at about $3~{\rm cm}^{-1}$ which
is assigned to magnetic correlations. \cite{Donovan97}
In full analogy to the case of UPd$_2$Al$_3$, the frequency 
dependence of the effective mass of UPt$_3$ displayed in 
Fig.~\ref{fig:UPt3}d clearly shows
that only a marginal increase of $m^*$ is observed around
30~cm$^{-1}$, while below the energies related to the magnetic correlations the 
mass is drastically enhanced.
Thus for  UPt$_3$ we also find that, as for UPd$_2$Al$_3$, the coupling of 
the localized and delocalized $5f$-electrons causes the heavy quasiparticles in  
UPt$_3$; these
magnetic excitations are very likely to be responsible for superconductivity. 
More detailed electrodynamic studies on UPt$_3$ and on the superconducting 
state in general have to be performed since they seem to be the appropriate 
method to probe the energy dependence of the correlation effects which are the 
key issue for understanding the nature of the heavy quasiparticles. 
It remains to be seen how far our findings have implications on non-magnetic 
heavy fermions with a large effective mass. Of further interest would be the study
of Ce compounds for which comparable band structure calculations do not yet exist.

\section{Conclusion}
In conclusion, the electrodynamic response of UPd$_2$Al$_3$ in the
low-energy range from 0.04 meV to 5 meV (0.33~cm$^{-1}$ to 40
cm$^{-1}$) exhibits a behavior at low temperatures ($T\leq 20~K$)
which cannot be explained within the simple picture of a
renormalized Fermi liquid. At around 12~meV a hybridization gap develops 
below $T^*$ which is characteristic for heavy fermions. 
The lower-frequency response shows the well-known
renormalized behavior only above the magnetic ordering; 
below $T_N=14$~K, however, additional features were discovered.
Besides an extremely narrow (less than
1~cm$^{-1}$) zero-frequency response, we observe a pseudogap of about 
0.22~meV. The
experiments yield indications that this pseudogap is not a simple
SDW gap or some excitation in the local magnetic moments system,
but rather is connected to magnetic correlations on the
delocalized charge carriers.
We have argued, that this gap at extremely low energies is due to the
influence of the localized magnetic moments on the itinerant electrons
and this interaction is mainly responsible for the enhanced mass $m^*$.
It appears that the formation of the heavy quasiparticles relies on the
establishment of antiferromagnetic order rather than competition of the coherent 
singlet formation and
the magnetic order.

\acknowledgements We acknowledge helpful discussions of our findings with 
F. Andres, A.A. Dolgov, A.V. Goltsev, G. Gr{\"u}ner, 
A. Muramatsu, A. Schwartz, F. Steglich, and G. Zwicknagl. 
We thank D.N. Basov for his attempt on the low-temperature
measurements. The work was partially supported by the Deutsche 
Forschungsgemeinschaft (DFG) via Dr228/9  and SFB~252.

\end{multicols}

\begin{thebibliography}{99}
\bibitem{remark5}An alternative explanation of the resistivity drop is based on the 
freezing out of the crystal field excitations.
\bibitem{remark1}The evaluation of effective charge carriers was done by using the 
relation: $(\omega_p^{\prime})^2 = \left.(\omega_p^*)^2\right|_{2{\rm K}} 
+8\left[\left. \int\sigma_1(\omega)\,{\rm d}\omega\right|_{300{\rm K}}\right. - \left.\left.\int\sigma_1(\omega)\,{\rm d}\omega \right|_{2{\rm K}}\right]$.
\end{thebibliography}
\end{document}